# SEMI-ACTIVE SUSPENSION CONTROL USING MODERN METHODOLOGY : A COMPREHENSIVE COMPARISON STUDY

**Omid Ghasemalizadeh[1], Saied Taheri[1], Amandeep Singh[2], Jill Goryca[2]**
[1]Mechanical Engineering Department, Virginia Polytechnic Institute & State University, Blacksburg, Virginia
[2]Analytics – Vehicle Dynamics & Durability Team U.S., Army RDECOM-TARDEC, Warren, Michigan

**Abstract**

*Semi-active suspensions have drawn particular attention due to their superior performance over the other types of suspensions. One of their advantages is that their damping coefficient can be controlled without the need for any external source of power. In this study, three control approaches are implemented on a quarter-car model using MATLAB/Simulink. The investigated control methodologies are Acceleration Driven Damper, Power Driven Damper, and $H_\infty$ Robust Control. The three controllers are known as comfort-oriented approaches. $H_\infty$ Robust Control is an advanced method that guarantees transient performance and rejects external disturbances. It is shown that $H_\infty$ with the proposed modification, has the best performance although its relatively high cost of computation could be potentially considered as a drawback.*

## I. Introduction

Human comfort and road holding of a vehicle are performance metrics that are greatly affected by the suspension system. Conventional (passive) suspensions are only efficient to some extent as they passively react to any disturbances introduced to the vehicle chassis by the road. Decades ago, active suspensions were proposed as an alternative to passive systems such as in [1]. Active suspensions were a step forward from passive suspensions; however, they needed an external source of power e.g. a hydraulic pump.

Several years later, semi-active suspensions were proposed where no external source of power was needed, except very small energy for running the electronics. The major difference between a semi-active suspension and a passive/active suspension is the use of a variable-damping concept where the damping coefficient of dampers can be controlled using an electronics unit. The damping ratio of semi-active suspensions is set in a closed-loop system with large bandwidth. Three main technologies have been developed for variable-damping dampers; electrohydraulic (EH), magnetorheological (MR), and electrorheological (ER).

One major concern of semi-active suspensions is developing an appropriate controller that determines the damping ratio needed for the best ride performance. Many control algorithms have been developed and implemented, both theoretically and practically. In 1994, Emura developed a controller based on the skyhook damper theory [2]. Later in 1999, Yi developed an observer-based control methodology [3] to estimate the velocity of sprung mass and unsprung mass by using the measured acceleration. The estimated error was independent of the unknown road disturbances. In 2000, Ahmadian experimentally tested the Skyhook, Groundhook, and the hybrid methods on a quarter-car rig with an MR damper [4]. He showed that the Skyhook method significantly reduces the sprung mass transmissibility while the Groundhook method improves the road holding characteristic.

In 2001, Yokoyama [5] developed a sliding mode control method, and used a target semi-active suspension as the model reference. His approach demonstrated reasonable robustness against model uncertainties and disturbances. Later in 2002, Choi [6] developed a more advanced method based on H∞ approach, and implemented it on a full-vehicle model. He treated the sprung mass as an uncertain parameter. In 2003, Sammer [7] compared H∞ with the well-known Skyhook theory by applying them on a nonlinear model. He concluded that as the design point of view H∞ showed improvement for both human comfort and road holding characteristic.

In 2008, Poussot-Vassal [8] introduced a new approach by implementing a linear parameter varying (LPV) on a nonlinear quarter-car model. Two main points of his work were a low computation cost of his method and a small number of sensors required. In 2012, Lozoya-Santos [9] compared LPV with a Frequency Estimation Based (FEB) principle. He showed that LPV performance can be modified by adjusting a set of matrices and FEB can be configured based on a look-up table electric current. Many different modern control schemes have been used for other applications such as $L_1$ adaptive control [10] or impedance control [11] that can be implemented on semi-active suspensions with some modifications.

In this study, two comfort-oriented approaches, namely, Acceleration Driven Damper (ADD), Power Driven Damper (PDD) were implemented on a linear quarter-car model. Ride performance metrics from the above approaches were compared to those from a modified H∞ Robust Control with a modification introduced later in this article. H∞ Robust Control is a modern control theory where robustness is guaranteed by canceling out the effect of external disturbances. Also, a passive suspension case was used for further analytical comparison.

The organization of the paper is as follows: the quarter-car model will be discussed in section II, then the three methodologies will be introduced and the theoretical assumptions behind each will be discussed in section III. In section IV, results of a set of simulations performed by MATLAB/Simulink on a quarter-car model as well as a 6-axle vehicle model will be shown and analytical comparison will be discussed. Finally, the conclusions of the study will be presented in section V.

## II. Problem Formulation

The quarter car model investigated in this study consists of a passive spring, a semi-active damper, the sprung and unsprung masses, and tire stiffness. The model is depicted in Figure 1.

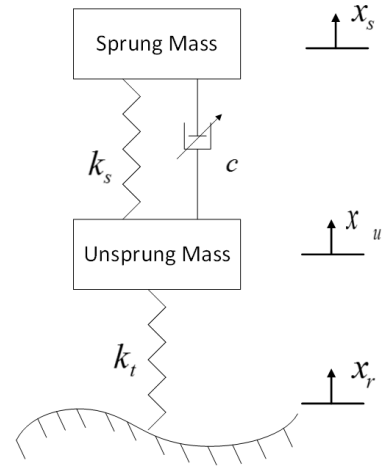

Figure 1 - Quarter-car model

The governing equations of motion for the quarter-car model are shown below:

$$m_s \ddot{x}_s + c \dot{x}_{def} + k_s x_{def} = 0 \quad (1)$$

$$m_u \ddot{x}_u - c \dot{x}_{def} - k_s x_{def} + k_t(x_u - x_r) = 0 \quad (2)$$

$$x_{def} = (x_s - x_u) \quad (3)$$

where $m_s$ and $m_u$ are sprung mass and unsprung mass, respectively. $c$, $k_s$, and $k_t$ are semi-active damper coefficient, suspension stiffness, and tire stiffness, respectively.

## III. Control Methodologies

In this section, three different approaches of controlling semi-active suspensions are briefly described and the mathematics behind each of them is shown. Further information regarding each methodology can be found in the references.

### A. Acceleration Driven Damper Control (ADD)

This approach was first introduced in [12]. This strategy is known as a simple yet effective control algorithm and shown to be optimal. It minimizes vertical acceleration of sprung mass by adjusting the damping coefficient. The ADD method sets damping coefficient as below:

$$c_{in} = \begin{cases} c_{min} & if\ \ddot{x}_s \dot{x}_{def} \leq 0 \\ c_{max} & if\ \ddot{x}_s \dot{x}_{def} > 0 \end{cases} \quad (4)$$

where $c_{in}$, $c_{min}$, and $c_{max}$ are the semi-active suspension damping coefficient, minimum damping coefficient and maximum damping coefficient, respectively. The suspension force $f$ is calculated using Equation (5):

$$f = c_{in} \dot{x}_{def} \quad (5)$$

This control approach is very well adapted for human comfort but the switching of damper coefficient values affects the closed-loop performance. In other words, ADD requires fast switching of damping coefficient, which is not as practical.

### B. Power Driven Damper Control (PDD)

PDD approach, introduced in [13], controls the energy stored and the power dissipated in a semi-active suspension. The results are comparable to those of ADD but the chattering effect of the control input is resolved to some extent. The proposed control law is shown in below equation:

$$c_{in} = \begin{cases} c_{min} & if\ k_s x_{def} \dot{x}_{def} + c_{min} \dot{x}_{def} \geq 0 \\ c_{max} & if\ k_s x_{def} \dot{x}_{def} + c_{min} \dot{x}_{def} < 0 \\ \dfrac{c_{min} + c_{max}}{2} & if\ x_{def} \neq 0\ and\ \dot{x}_{def} = 0 \\ -\dfrac{k_s x_{def}}{\dot{x}_{def}} & Otherwise \end{cases} \quad (6)$$

Again, the suspension force is calculated from (5). The advantage of this approach is its low-chattering performance although it needs knowledge of suspension stiffness value. Switching of damping coefficient is not too fast as in ADD.

### C. H∞ Robust Control

H∞ algorithm guarantees stabilization and robustness by modeling the system as an optimization problem. It is considered as a modern control technique that was developed in late 1970s – early 1980s. The name comes from the mathematical H∞ norm that represents the maximum singular value of a matrix function in Laplace space that is bounded in the right-half plane. The only drawback with this method is handling of non-linear constraints such as saturation. The controller formulation will be briefly discussed in next few paragraphs.

Assume the closed loop system shown in Figure 2, where $P$ is the plant, $K$ is the controller feedback, $w$ is the external disturbance, $z$ is the variable to be minimized, $y$ is the plant output, and $u$ is control input. It should be noted that $w$, $z$, $y$, and $u$ are vectors but $P$ and $K$ are matrices.

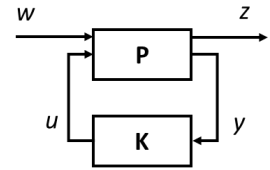

Figure 2 - H∞ closed loop system

In system showed above, the goal is minimizing of the error variable ($z$). Algebraic representation of the closed loop model is as following:

$$\begin{bmatrix} z \\ y \end{bmatrix} = P(s) \begin{bmatrix} w \\ u \end{bmatrix} \quad (7)$$

and from Figure 2 it can be interpreted that $u = K(s)y$. Therefore, the transfer function $F_l(P, K)$ from exogenous input $w$ to minimized output $z$ can be written as

$$z = F_l(P, K)w \quad (8)$$

The objective of H∞ algorithm is to find a feedback controller $K$ that minimizes the H∞ norm of $F_l(P, K)$ [14]. As a side note, the idea is very similar to H2 control design.

In this study the plant $P$ is a quarter-car model equipped with a semi-active suspension as described in section II. Also, $w$ is the road profile and $u$ is the control output which is the damping coefficient for the semi-active suspension. In reality, the damping value is bounded between two positive minimum and maximum values. In other words:

$$0 < c_{min} \ll c \ll c_{max} \quad (9)$$

where $c_{min}$ and $c_{max}$ are the minimum and maximum values that the semi-active damper can acquire, respectively.

According to equation (9), a saturation operator is required to clip the calculated damping value and feed it into the acceptable range. Thus, the modified schematic block diagram becomes as shown in figure below:

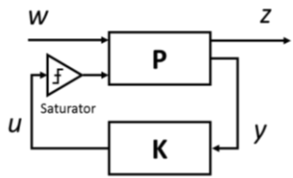

Figure 3 - Modified H∞ algorithm block diagram

As mentioned previously, H∞ cannot guarantee stability in presence of a saturation constraint. Therefore, a tweak to the above control scheme is introduced to resolve the instability issue. The proposed modification assumes a mathematical model similar to the main quarter-car model but without any constraints, and implements the H∞ control on that model. On the side, the main model is fed with the clipped control input that comes from the controller. The proposed H∞ control scheme is shown in Figure 4 below.

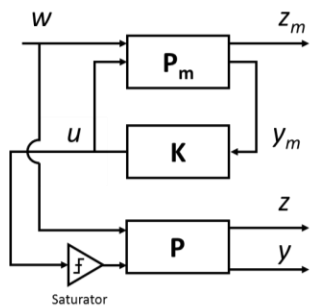

Figure 4 - Proposed modification for H∞ Control

In above control method, a mathematical model of the quarter-car without a saturation constraint for the semi-active damper ($P_m$) is controlled by feedback control according to H∞ control algorithm. Furthermore, the control input is saturated and fed to the main plant $P$ (with saturation constraint). Using introduced method, the system does not involve any instability issue although the performance of the main plant with saturation constraint is compromised. It is a sacrifice of performance to resolve singularity issue of the closed loop system.

## IV. Results and Discussions

### A. Quarter-car simulation

The control algorithms discussed above was implemented on a MATLAB/Simulink quarter-car model. The quarter-car parameters were selected to represent a heavy truck. The sprung mass was assumed to be $m_s = 2250 kg$, unsprung mass was assumed to be $m_u = 200 kg$, the minimum damping coefficient was $c = 2000 \, N \cdot s/m$, the maximum damping coefficient was $c = 40000 \, N \cdot s/m$, suspension stiffness was $k_s = 180000 \, N/m$, and the tire stiffness assumed to be $k_t = 500000 \, N/m$. Also, a passive system was considered for comparing the semi-active suspension algorithms with a passive suspension case. The passive damper value was assumed $c = 5000 \, N \cdot s/m$. Also, the road-profile shown in Figure 5 was used.

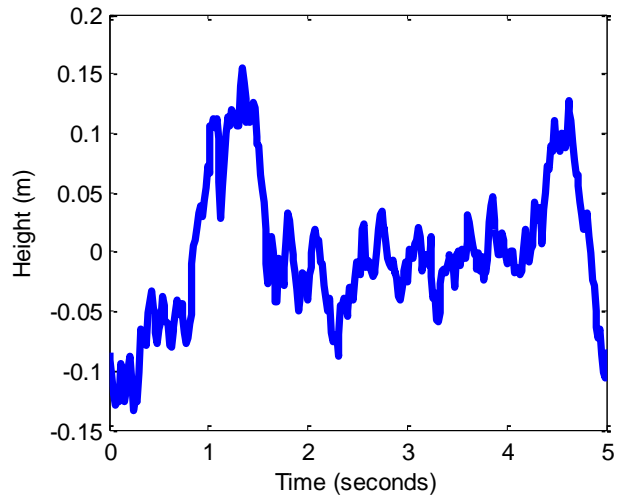

Figure 5 - Road profile

The quarter-car simulation model was developed using MATLAB/Simulink and control algorithms were implemented using the same tool. Figure 6 shows the sprung mass displacement vs. time for the passive system and the three algorithms discussed above.

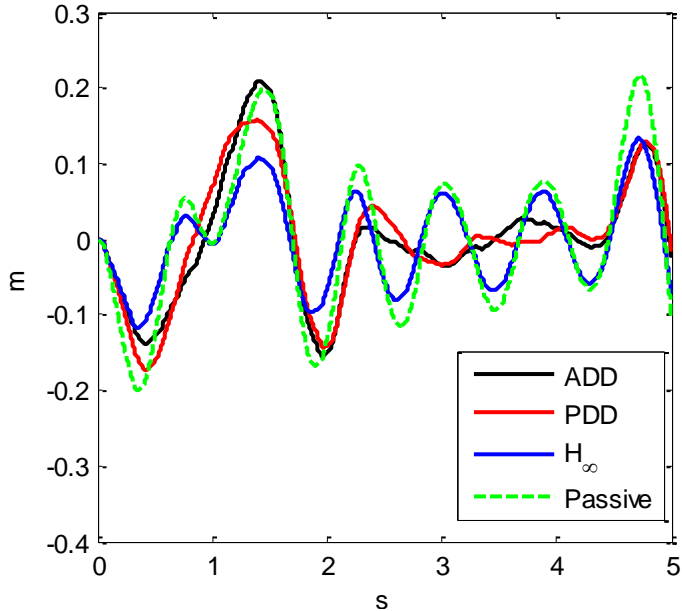

Figure 6 – Sprung mass displacement for different control approaches.

As shown in Figure 6, PDD and ADD methods keep almost the same curve, but PDD has smaller magnitudes in some times. Moreover, $H_\infty$ shows lower magnitude of sprung mass displacement compared to the other three. PDD and ADD show some chattering effect that has a negative impact on vertical acceleration of sprung mass. Note that, as mentioned before, PDD reduces chattering effect in control input (i.e. damping coefficient), not sprung mass displacement.

Figure 7 shows the acceleration plot of each of the cases. The plots are separated for better visual display. Passive case has the highest acceleration magnitude as expected. $H_\infty$ has the lowest perturbations of acceleration. ADD and PDD are similar but the advantage of PDD is having zero acceleration in some time intervals. This can be realized from the PDD control law equation (6). When the *otherwise* condition holds, the spring force applied on the sprung mass is neutralized by the damper force. In other words, the total amount of force applied on the sprung mass is zero and consequently, the acceleration becomes zero as well.

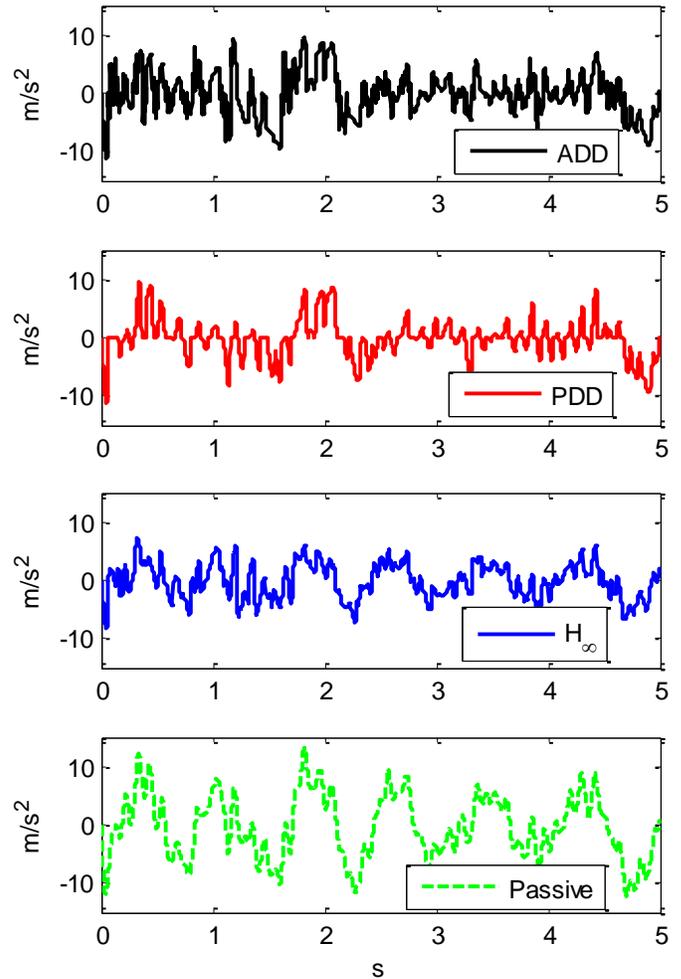

Figure 7 - Sprung mass acceleration for different control algorithms

To study performance of each method the absorbed power of each case was calculated over the 5 seconds simulation time. As shown in Figure 8, passive case builds up the highest absorbed power. ADD and PDD are showing similar behavior up to 1s and after that PDD is settling down to a lower value. Also, $H_\infty$ has the lowest absorbed power compared to the other cases.

To investigate human comfort metric of each approach, quantified values are required along with the shown graphs. For acceleration of sprung mass, root mean square (RMS) values of vertical acceleration is widely used in the literature as the human comfort metric. Absorbed power is another quantity that expresses human comfort metric. In this study, a procedure was used that averages the absorbed power from the initial moment till present. Thus, the final absorbed power, over the course of 5 seconds simulation, can be realized better from the final data points of graphs in Figure 8. In order to have a quantified value of the absorbed power, the average of its values over the last 0.5s of simulation was calculated and shown in Table 1.

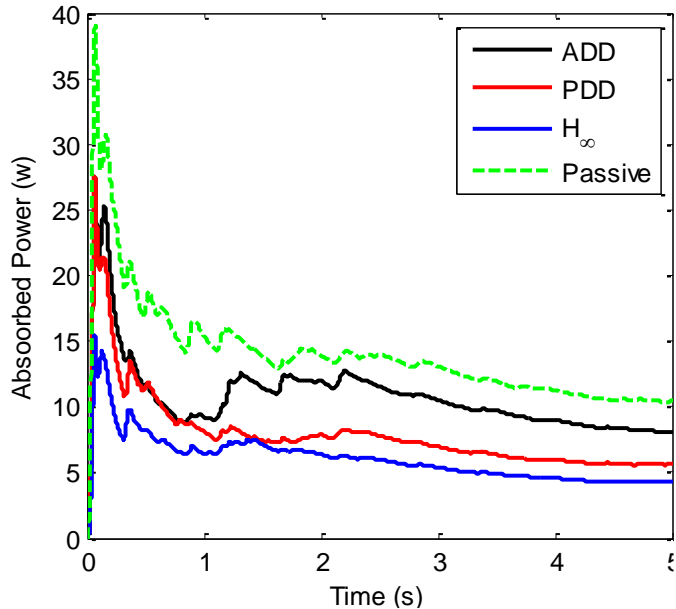

Figure 8 - Absorbed power of different control algorithms

According to the results summarized in Table 1, $H_\infty$ is by far the best approach. Roughly speaking, it out performs the second best method, PDD, by a 10% improvement on sprung mass acceleration RMS and 23% improvement on absorbed power. Note that the absorbed power mentioned in the table, is the average of values shown in the Figure 8 over last 0.5s of the simulation.

As a modern method, $H_\infty$ needs more computational time and its implementation is more complicated compared to the other approaches. Note that this simulation has been performed only for a quarter-car model. The computation load will be much more for a full-vehicle where at least 4 suspensions are to be controlled simultaneously.

ADD has worse RMS value of sprung mass acceleration than PDD. This fact also can be realized from the sprung mass displacement curve, Figure 6.

The corresponding curve to ADD has slightly more chattering effect and that leads to higher RMS value of acceleration, which is the second derivative of displacement with respect to time.

Table 1 – RMS values of sprung mass acceleration and absorbed power (average of over last 0.5s)

| Method | RMS($\ddot{x}_s$) | Absorbed Power |
|---|---|---|
| ADD | 3.764 | 8.2293 |
| PDD | 3.3885 | 5.6376 |
| $H_\infty$ | 3.0487 | 4.3078 |
| Passive | 5.5165 | 10.4842 |

PDD performs better than ADD and its low-chattering control input is considered as an advantage over ADD. In real-time applications, too much chattering in input signal cannot be performed by the actuators with relatively smaller bandwidths.

**B. 6-axle vehicle simulation**

In section IV.A a quarter-car model was used to implement the aforementioned approaches. To compare the capability of those methods, further investigation on a full-car model is required. ADD and PDD are corner independent approaches but on the other hand, $H_\infty$ controls the whole model as one system. In other words, for a full-vehicle model with $n$ corners/suspensions, $n$ ADD/PDD controllers are needed whereas only one $H_\infty$ controller is required to control the ride quality. Saying above, pitch and roll angles are taken care of implicitly in $H_\infty$ but it is not the case for ADD/PDD.

In this section, results of implementing the controllers on a 6-axle car model (12 corners/suspensions) will be presented. A simulation in MATLAB/Simulink environment was performed using below parameters. The vehicle assumed to have 9000kg of sprung mass ($m_s = 9000 kg$), each unsprung mass is $m_u = 200 kg$, the minimum damping coefficient was $c = 2000 \, N \cdot s/m$, the minimum damping coefficient was $c = 40000 \, N \cdot s/m$, suspension stiffness for the front axle was $k_s = 130000 \, N/m$ and it was $k_s = 180000 \, N/m$ for the rest of the axles, and the tire stiffness assumed to be $k_t = 500000 \, N/m$ for all the axles. The distances between the second, third, fourth, fifth, and sixth axles to the front axle are 1, 1.9, 2.4, 3, and 3.6 meters, respectively. Also, the center of gravity is located 1.7m away from the front axle. The track width is 2m and the center of gravity is assumed to be at the middle of the vehicle i.e. 1m away from both left and right suspensions. A schematic view of the vehicle is shown in the appendix.

Below road profiles (left and right wheels road profiles) were used in the simulation.

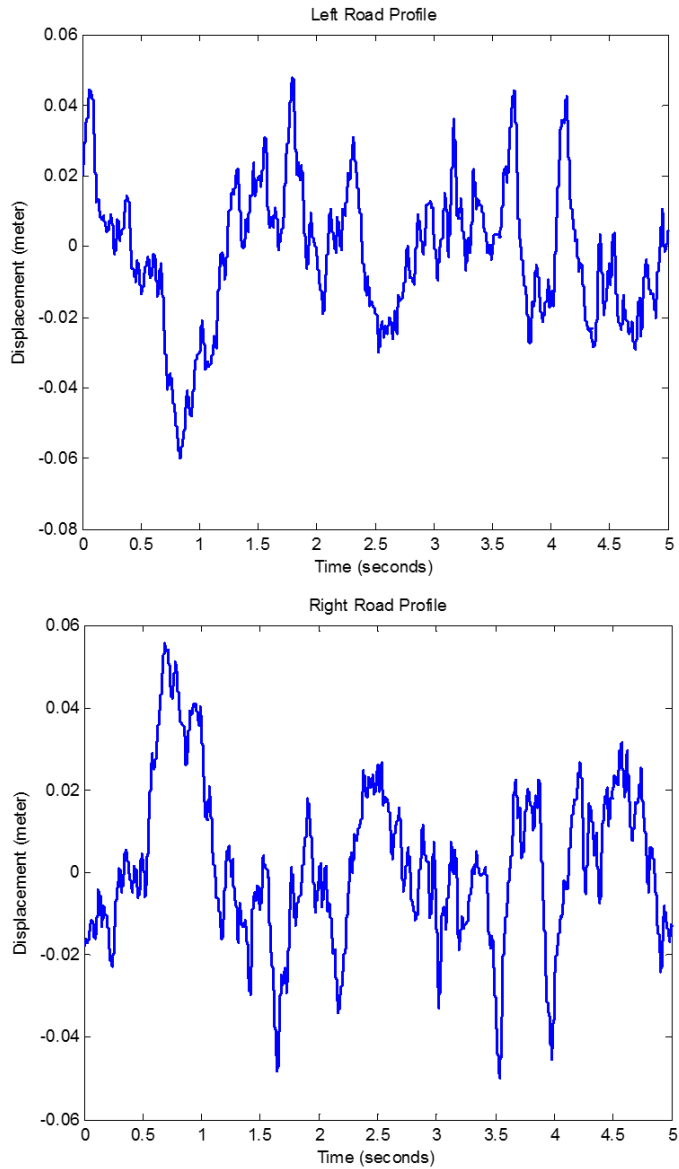

Figure 9 - Left and right road profiles used for the 6-axle vehicle model simulation

The three controllers were implemented on the 6-axle model and a set of simulations were run to compare their performances. Figure 6 show the center of gravity displacement.

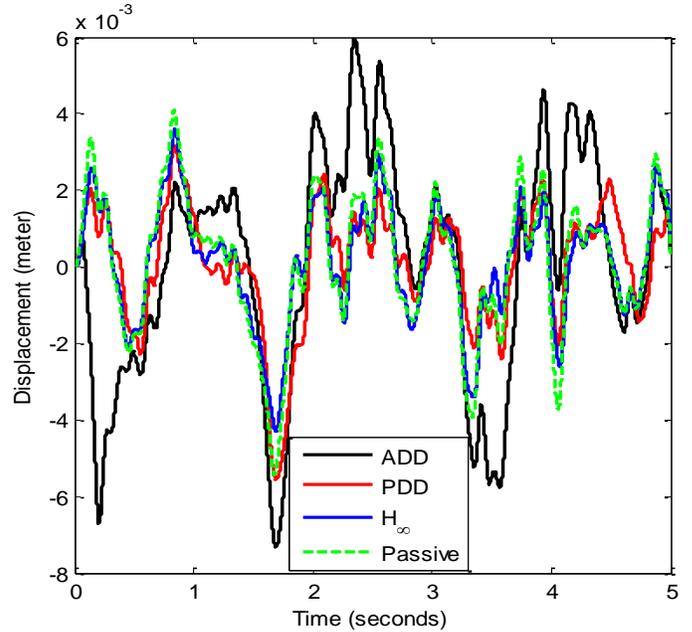

Figure 10 - 6-axle vehicle center of gravity displacement

Visually speaking, ADD has the highest amplitude of center of gravity displacement, PDD and $H_\infty$ have lowest amplitudes and passive is in between. Since center of gravity displacement does not show the ride quality, acceleration plots are needed for further investigations.

Acceleration plots are shown in Figure 11. Again, PDD and $H_\infty$ have lowest acceleration amplitudes and ADD has the highest spikes. In other words, human comfort index (acceleration RMS) is expected to be highest for ADD. Passive suspension is working better than ADD in the 6-axle model and that is because ADD controls each corner independently and does not take care of the roll and pitch angles. PDD controls each corner separately too but based on the acceleration plots, its control law is good enough for a full-car model. Accurate comparison of PDD and $H_\infty$ is not possible yet as more information such as absorbed power is needed.

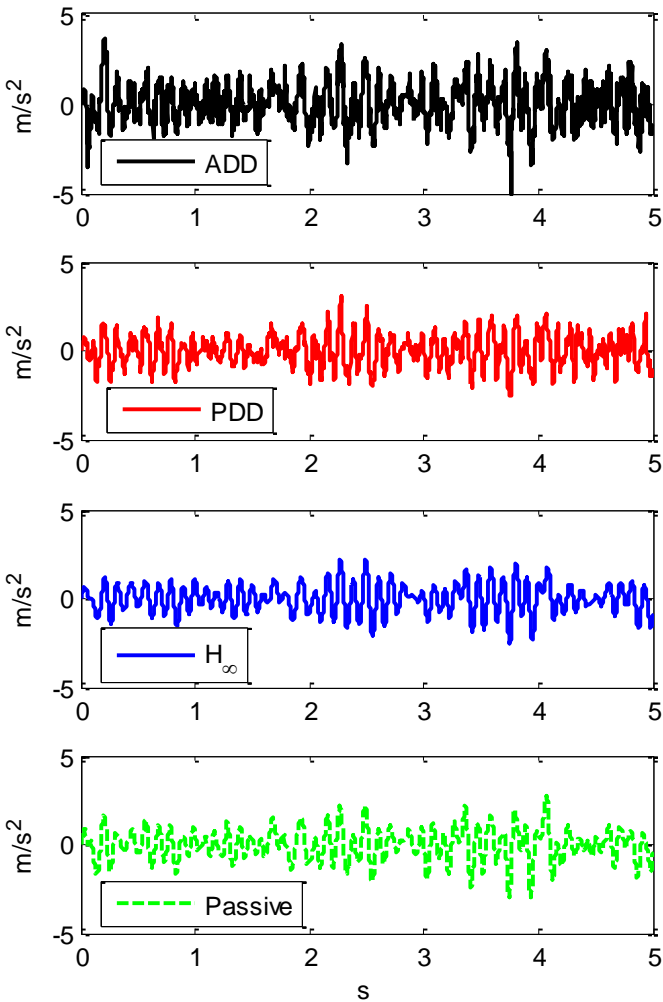

Figure 11- Center of gravity acceleration for 6-axle vehicle model

Figure 12 shows the absorbed power plots for the controllers. As seen, ADD builds up the highest amount of absorbed power over the 5-second simulation. $H_\infty$ has best performance in terms of absorbed power. Moreover, PDD is slightly worse than $H_\infty$ and passive has the third performance and is better than ADD.

As discussed previously, since ADD is a corner independent method, it only controls the acceleration of the point that it is connected to the sprung mass. It is a drawback of that concept because dynamics of all the corners are interconnected to each other by pitch and roll effects and performing independently from the other corners reduces chance of performing as well as expected. PDD has the same issue but apparently its concept of reducing the absorbed power as mentioned in III.B increases its efficiency.

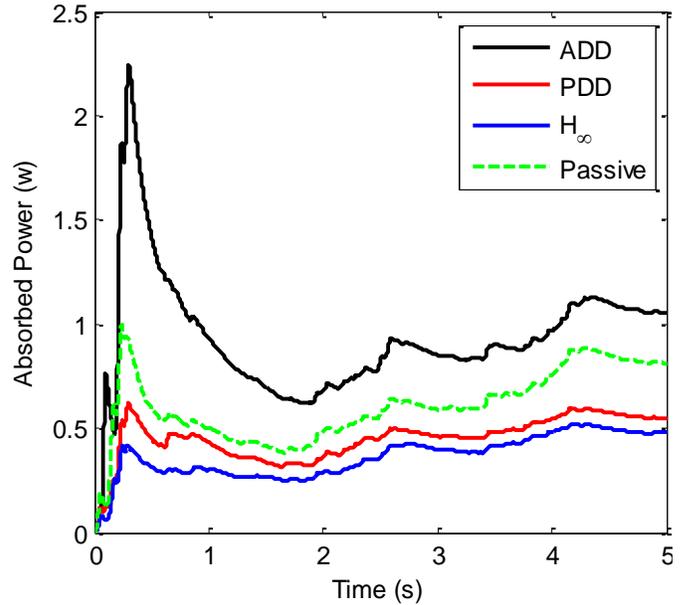

Figure 12 - Absorbed power for 6-axle vehicle model

The same as quarter-car model, a table of quantified values of center of gravity acceleration and absorbed power plots provided below. Once again, the absorbed power value in the table is the average of last 0.5s values from the graph.

Table 2 - RMS values of sprung mass acceleration and absorbed power (average of over last 0.5s)

| Method | RMS($\ddot{x}_s$) | Absorbed Power |
|---|---|---|
| ADD | 1.1208 | 1.0701 |
| PDD | 0.8836 | 0.5585 |
| $H_\infty$ | 0.8319 | 0.4841 |
| Passive | 0.9212 | 0.8246 |

In this case, $H_\infty$ has improved the RMS of acceleration by 6% compared to the second best approach PDD. Moreover, the absorbed power of $H_\infty$ has been improved by 13% with respect to PDD.

Studying of vehicle corners, and roll and pitch effects can be performed as the future of works of this investigation. Also, using different road profiles would give a better understanding of how each method works.

## V. Conclusions

First, a quarter-car model with a semi-active suspension was introduced. In this model a passive spring along with a semi-active damper was used as show in Figure 1.

Three control algorithms, namely, ADD, PDD, and $H_\infty$ were investigated in this paper. The control methodologies of ADD, and PDD, which are known as comfort-oriented methodologies, were directly taken from their respective references. The modified $H_\infty$ Robust Control algorithm was developed during the course of the research. $H_\infty$ is known as a modern control scheme where the focus is on canceling out the effect of external disturbances and the robustness of closed loop system.

MATLAB/Simulink was used to perform a set of simulations on the developed quarter-car model. Results of the simulations showed that modified $H_\infty$ has the best performance. RMS values of sprung mass acceleration and absorbed power were used as the comparison metrics. The only drawback with $H_\infty$ is its relatively high cost of computation.

$H_\infty$ showed the best performance based on the human comfort metrics, RMS of sprung mass acceleration and absorbed power. It showed 10% improvement on sprung mass acceleration RMS and 23% improvement on absorbed power when compared to PDD, the second best approach. PDD had the best performance amongst the simple comfort-oriented approaches. ADD showed slightly worse performance than PDD. As mentioned in section IVIII.B, PDD reduces the effects of chattering in control input as compared to ADD
. Existence of chattering effect in control input is considered as a disadvantage of ADD compared to PDD.

For studying effects of roll and pitch angles on performance of each of the methodologies, a 6-axle vehicle model was developed. The developed vehicle model is depicted in the appendix. A simulation using MATLAB/Simulink tool was performed with a set of given vehicle parameters.

The simulation showed that $H_\infty$ improves ride quality by reducing the center of gravity acceleration RMS as well as center of gravity absorbed power for the given road and vehicle data. $H_\infty$ showed 6% enhancement on center of gravity acceleration RMS and 13% improvement on absorbed power compared to PDD, which was the second best approach. ADD had the lowest performance metrics even compared to a passive case. Because ADD is a corner independent method that only tries to minimize the acceleration of the point that it is attached to the sprung mass without having any information of other corners. The fact that the corners dynamics are interconnected by roll and pitch effects, reduces the ADD performance. PDD is a corner independent approach too but based on the obtained results, its concept of reducing the absorbed energy allows it to perform better than ADD and passive.

## VI. Future Works

As mentioned earlier, a more detailed study on the effects of roll and pitch angles can be done especially on the corners of the vehicle where the suspensions are connected to the sprung mass. Also, a modified control algorithm can be developed that optimizes the controller performance based on the seats positions as improvement of ride quality at those points is highly desired.

Also, a quarter-car suspension test rig has been designed and fabricated in the Center for Tire Research (CenTiRe) lab. This facility can be used for validation of simulation results of this study and similar ones. An in-detail investigation can be performed to compare the simulation result to those of obtained from the experimental test rig.

The next phase of this project is defined as developing other control algorithms and validation of simulation results using the experimental suspensions rig.

Appendix

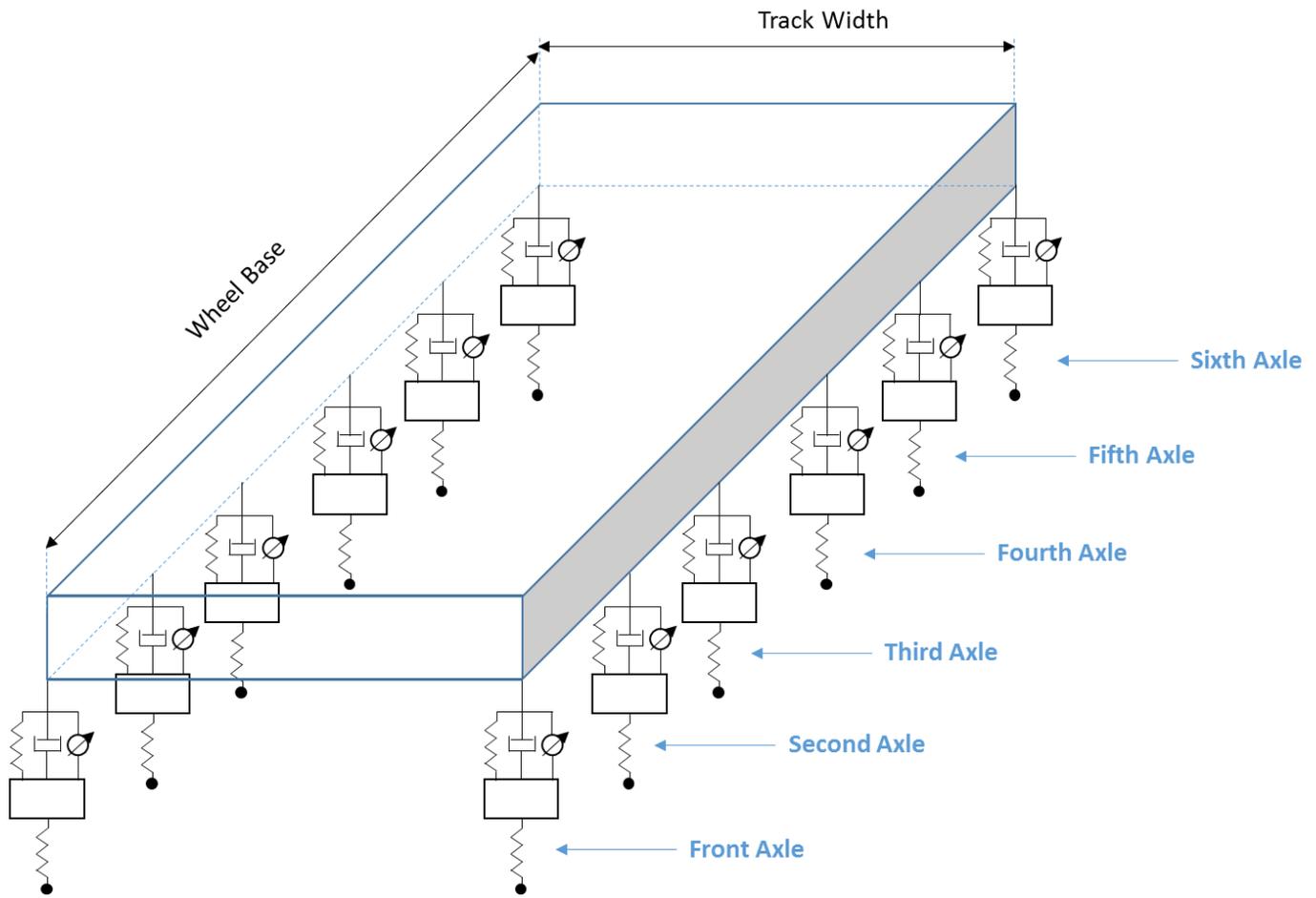

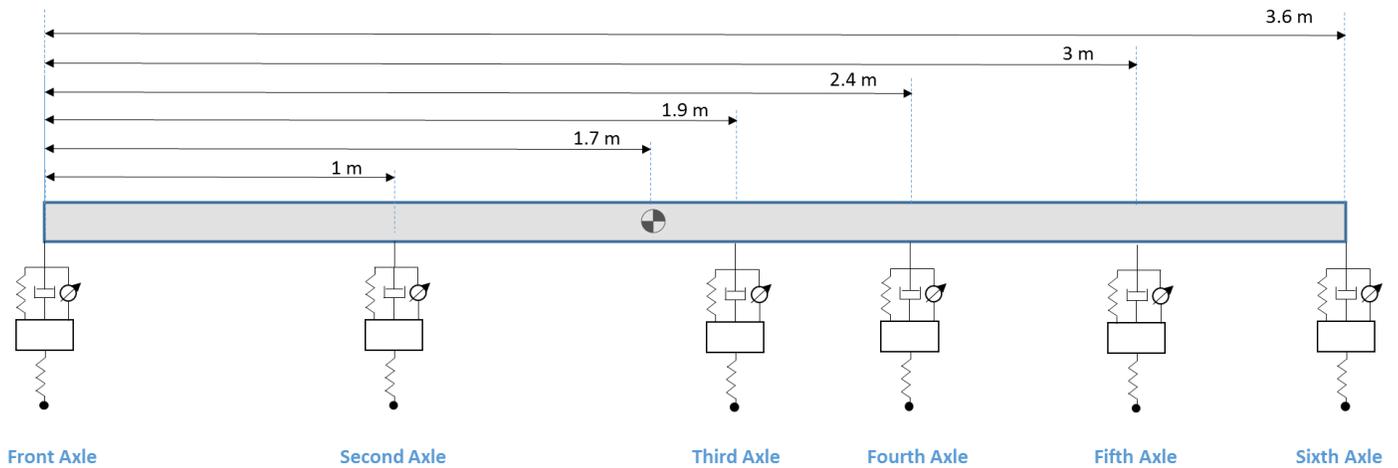